\documentclass[letter]{aa}
\usepackage{graphicx}
\usepackage{txfonts}
\usepackage{natbib}
\def\HII{{\ion{H}{II}}}
\def\HI{{\ion{H}{I}}}
\def\OIII{{[\ion{O}{III}]}}
\def\SII{{[\ion{S}{II}]}}
\def\NII{{[\ion{N}{II}]}}
\def\OI{{[\ion{O}{I}]}}
\newcommand{\kms}{$\,$km$\,$s$^{-1}$}

\begin{document}

\title{The outer filament of Centaurus A as seen by MUSE}
\titlerunning{MUSE observations of Cen A}
\author{F. Santoro \inst{1,2}\fnmsep\thanks{email: santoro@astro.rug.nl},
             J. B. R. Oonk\inst{1,3},
             R. Morganti\inst{1,2},
             T.A. Oosterloo\inst{1,2},
             and G. Tremblay\inst{4} }

\institute{ASTRON, Netherlands Institute for Radio Astronomy, PO 2, 7990 AA, Dwingeloo, Netherlands.\and Kapteyn Astronomical Institute, University of Groningen, PO 800, 9700 AV Groningen, Netherlands.\and Leiden Observatory, Leiden University, PO Box 9513, 2300 RA Leiden, the Netherlands. \and Yale University, Department of Physics, 217 Prospect St., New Haven, CT 06511, USA}

\date{Received 12/12/2014; accepted 23/01/2015}
 
\abstract {Radio-loud active galactic nuclei (AGN) are known to inject kinetic energy into the surrounding interstellar medium of their host galaxy via plasma jets. 
Understanding the impact that these flows can have on the host galaxy helps to characterize a crucial phase in their evolution.  Because of its proximity, Centaurus A is an excellent laboratory in which the physics of the coupling of jet mechanical energy to the surrounding medium may be investigated.  About 15 kpc northeast of this galaxy, a particularly complex region is found: the so-called outer filament, where jet-cloud interactions have been proposed to occur. }
{ We investigate signatures of a  jet-interstellar medium (ISM)  interaction using optical integral-field observations of this region, expanding on previous results that were obtained on a more limited area.}
{ Using the Multi Unit Spectroscopic Explorer (MUSE) on the VLT during the science verification period, we observed two regions that together cover a significant fraction of the brighter emitting gas across the outer filament. Emission from a number of lines, among which H$\beta~\lambda$4861\AA, \OIII$\lambda\lambda$4959,5007\AA, H$\alpha~\lambda$6563\AA, and \NII$\lambda\lambda$6548,6584\AA,\ is detected in both regions.}
{ The ionized gas shows a complex morphology with compact blobs, arc-like structures, and diffuse emission. Based on the kinematics, we identified three main components of ionized gas.  Interestingly, their morphology is very different.  The more collimated component is oriented along the direction of the radio jet.  
The other two components exhibit a diffuse morphology together with arc-like structures, which are also oriented along the radio jet direction. Furthermore, the ionization level of the gas, as traced by the \OIII$\lambda$5007/H$\beta$ ratio, is found to decrease from the more collimated component to the more diffuse components. }
{ The morphology and velocities of the more collimated component confirm the results of our previous study, which was limited to a smaller area, implying that both the outer filament and the nearby \HI\ cloud are probably partially shaped by the lateral expansion of the jet. The arc-like structures embedded within the two remaining components are the clearest evidence of a smooth jet-ISM interaction along the jet direction.
We thus find signs of a jet-ISM interaction across all identified gas components. This suggests that, although poorly collimated, the large-scale radio jet is still active and affects the surrounding gas. This result indicates that the effect on the ISM of even low-power radio jets should be considered when studying the influence AGN can have on their host galaxy.
 
       }
   

\maketitle
%

\section{Introduction}

Active galactic nuclei (AGN) can influence the evolution of their host galaxy by injecting energy into the surrounding interstellar medium (ISM) \citep[see][for a review]{2014ARA&A..52..589H}. The mechanical energy provided by a radio-loud AGN is mainly in the form of twin jets launched into the ISM, which can represent a key ingredient in the overall evolution of an active galaxy.  Indeed, the interaction between radio jets and ISM has been found to be relevant at large and small scales both by observations \citep[see][and references therein]{2012NJPh...14e5023M,2013Sci...341.1082M} and numerical simulations \citep[e.g.,][]{2012MNRAS.425..438G,2012ApJ...757..136W}.

How the interaction between the jet and the ISM proceeds, how it depends on the kinetic power of the jet, for instance, and whether this process is still strong in the more common low-power radio galaxies, are some of the open questions. Several gas outflows are known that are driven by low-power radio jets, including cases of massive neutral and molecular outflows \citep[e.g.,][]{2011ApJ...735...88A,2013A&A...558A.124C}. 
  This indicates the possibility that even jets with low kinetic power can drive outflows, suggesting that the jet-ISM interaction could be a relatively common phenomenon associated with the active phase of a galaxy \citep[see also][ for a discussion]{2014MNRAS.441.3306H}.

  Detailed observations are a crucial starting point for understanding the physics of jet-ISM interactions. Furthermore, they can place important constraints on the main parameters of the models used
to simulate these interactions.  Signatures of jet-ISM interactions have also been found in the nearby AGN Centaurus A \citep[Cen~A, see][]{2002ApJ...564..688R,2005A&A...429..469O,2014arXiv1409.7700H,2014arXiv1411.4639S}. This led us to choose this galaxy as a target for a detailed study of the jet-ISM interaction phenomenon.  Cen~A is a relatively low-power radio galaxy, \citep[$\rm{P_{2.7GHz}=1.8\times 10^{24}W Hz^{-1}}$;][]{1999MNRAS.307..750M} as well as the nearest observed AGN \citep[$d=3.8$ Mpc;][for which 1 arcmin =1 kpc]{2010PASA...27..457H}, with a systemic velocity v$_{\rm sys}\sim$540 \kms.  Filaments of highly ionized gas have been found at different locations along the jet direction \citep{1975ApJ...198L..63B,1981ApJ...247..813G,1991MNRAS.249...91M}. Past studies mainly focused on the  so-called inner and outer filaments situated at a distance of about 8.5 kpc and 15 kpc from the galaxy nucleus. In both cases, the radio plasma jet is thought to affect the characteristics of the gas.  Indeed, these filaments are close to the inner radio lobe and to the large-scale jet \citep[see e.g. discussion in][]{1999MNRAS.307..750M}.

The outer filament region presents a particularly intriguing situation, where a complex interplay between radio plasma and neutral and ionized gas is ongoing. Interestingly, only a low surface brightness and poorly collimated jet-like structure has been observed in this region \citep{1999MNRAS.307..750M}. Whether this structure corresponds to a still active jet or to a relic is still a matter of discussion \citep[e.g.,][]{2001ApJ...563..103S, 2009MNRAS.393.1041H}. 
Indirect evidence of interaction - with associated star formation - at this location has been found by a number of studies \citep{2000ApJ...536..266M,2002ApJ...564..688R,2005A&A...429..469O}.
\citet{2005A&A...429..469O} studied the kinematics of the  \HI\ cloud, located - in projection - close to the outer filament and found
that gas with anomalous velocities is a possible interaction
signature.

\citet{2014arXiv1411.4639S}  carried out an integral-field spectroscopic study of the ionized gas using the Visible MultiObject Spectrograph (VIMOS) on two small regions of the outer filament. They found two kinematical components whose velocities match those of the two components found in the nearby \HI\ cloud. These data support the idea that ionized and neutral gas are likely part of a single dynamical structure that is partly affected by the interaction with a poorly collimated, but active jet.  If this is  the case, the kinematical signature of the jet-ISM interaction traced by \citet{2014arXiv1411.4639S} should be seen over a larger area of the outer filament. With the goal of extending our study over a larger region, we have exploited the new capabilities provided by the Multi Unit Spectroscopic Explorer \citep[MUSE,][]{2010SPIE.7735E..08B} integral field spectrograph that was recently installed at the ESO/Very Large Telescope (VLT). 
In this letter we show the first results on the kinematics of the gas, while a detailed analysis of the ionization and line ratios will be presented in a forthcoming paper.
 
\begin{figure*}[t]
\centering
\includegraphics[width=16 cm, keepaspectratio]{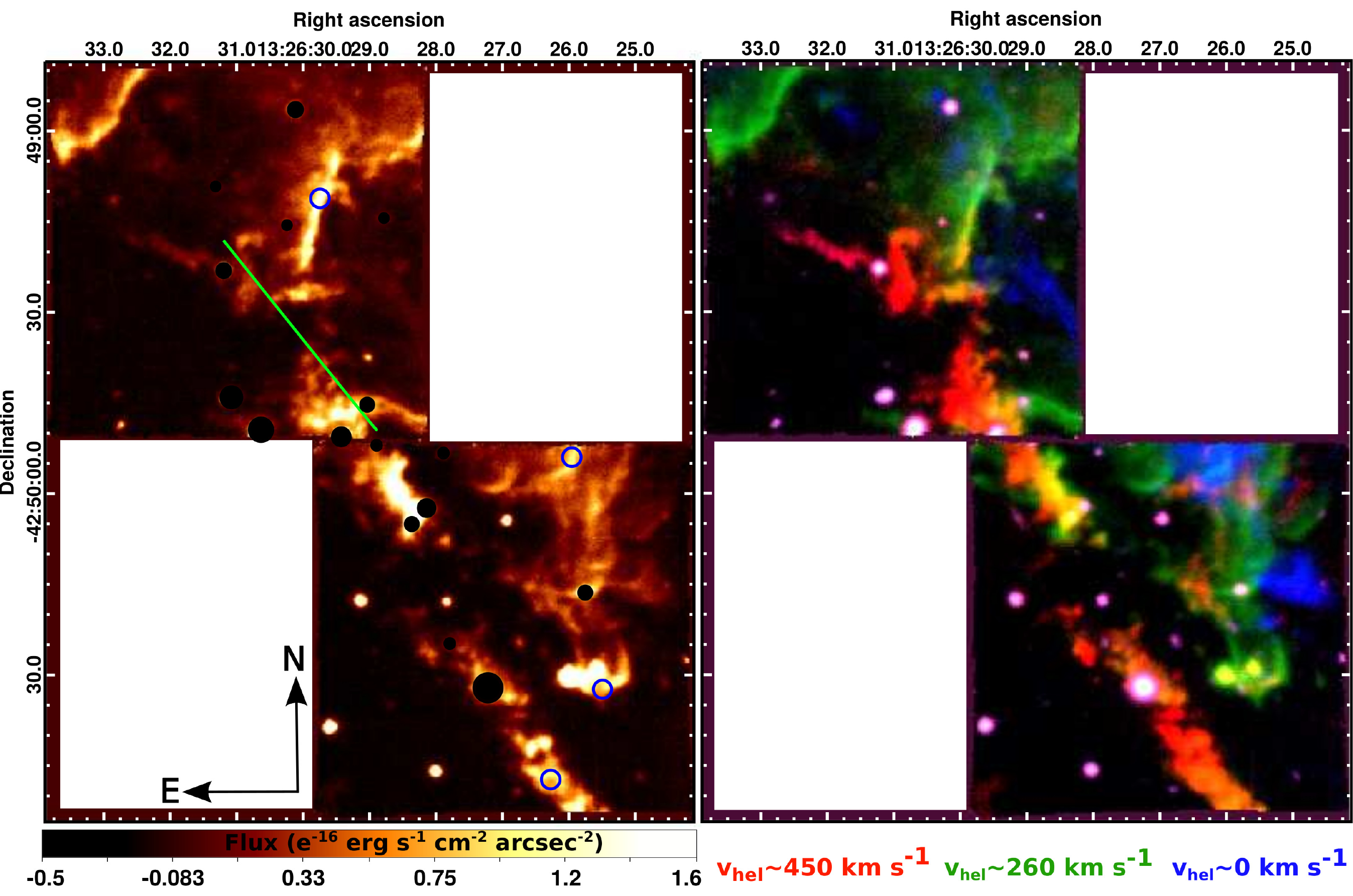}
\caption{\textit{Left panel}. Total H$\alpha$ line flux map. Blue circles mark the regions from which the spectra shown in Fig.~\ref{Spectra} are extracted (from top to bottom: diffuse component, low-velocity component, arc-like clumps, linear component). Black-filled circles mark the stellar light that is superimposed on the ionized gas emission. The location of the position-velocity plot  of Fig.~\ref{pvplot} is indicated with a green line. \textit{Right panel}. RGB image of the H$\alpha$ emission showing the linear (red), the diffuse (green) and the low-velocity (blue) components. At the bottom we indicate the intensity weighted mean heliocentric velocity of each component; note that Cen~A v$_{\rm sys}\sim$540 \kms. The regions where the overlap between the linear and the diffuse components is more evident are plotted in yellow.}
\label{TotalFluxMap}
\end{figure*}

\section{Data reduction and analysis}
We observed two fields in the outer filament of Cen~A using MUSE at the VLT.
Observations were carried out on June 25, 2014 during the science verification program with an exposure time of 3$\times$16.6 min for each field.
Our two MUSE pointings cover the brighter region of the outer filament, which partially overlaps the our previous VIMOS observations \citep{2014arXiv1411.4639S}. Compared to that study, we increased the observed area from $\sim$0.5 to 2 arcmin$^2$ and now cover a large part of the outer filament. Furthermore, we extend the observed wavelength range beyond the 6200 \AA\ VIMOS limit to 9300 \AA\ with MUSE. This enables us to include the important diagnostic lines from the red part of the optical spectrum.

The observations were made in Wide Field Mode  with a field of view (FoV) covering 1$\times$1 arcmin$^{2}$ and a pixel size of $0.2 \times 0.2$ arcsec$^{2}$. The spatial resolution is limited by the seeing, which was $\sim$1 arcsec. The spectral range is about 4700-9300\AA\ and the spectral resolution is  $\sim 2.3$\AA,\, corresponding to a velocity resolution ranging from 75 \kms\ at the longest wavelengths to 150 \kms\ at the shortest one. For flux calibration purposes we used the standard calibration procedure. The data reduction was carried out using the ESO pipeline (Version 0.18.2) with default parameters \citep{2014ASPC..485..451W}. 
The sky subtraction process left residuals at the location of bright sky lines, but these do not compromise the quality of the science discussed here.
Emission lines from ionized gas are clearly detected over the whole wavelength range and across both fields and include H$\beta~\lambda$4861\AA, \OIII $\lambda\lambda$4959,5007\AA, H$\alpha~\lambda$6563\AA, \NII $\lambda\lambda$6548,6584\AA, \SII$\lambda\lambda$6717,6731\AA, and \OI$\lambda\lambda$6300,6363\AA. All the lines show the same spatial and velocity distribution.

\section{Results}

The integrated emission of H$\alpha$ given in Fig.~\ref{TotalFluxMap} (left panel) shows an intriguing variety of gas structures spread over the entire observed region. The gas distribution we observe across our FoV is consistent with the previous optical studies of the outer filament \citep{1991MNRAS.249...91M,1998ApJ...502..245G,2001A&A...379..781R,2014arXiv1411.4639S}, but we obtain for the first time detailed kinematical information over a large area of the filament. The gas is distributed in structures that range from compact and bright to faint, extended,  and diffuse. Particularly interesting are the well-defined arc-like features in the diffuse emission.
The MUSE data allow us to distinguish three kinematical components, which are represented in Fig. 1 (right panel).  This figure shows an RGB image of the H$\alpha$ emission colour-coded based on the heliocentric velocity of the line emission.  Here we have separated the components by summing the line emission in three heliocentric velocity ranges: $-$108-177 \kms\ (blue), 177-405 \kms\ (green), 405-745 \kms\ (red). 
 All these structures are blueshifted relative to the systemic velocity of Cen~A.  Two components (shown in green and red) correspond - extending over a larger area - to those reported by \citet{2014arXiv1411.4639S}. However, our new data reveal a third, lower velocity component (shown in blue). Because of the limited FoV, it was not detected in the VIMOS study. Hints of this component were previously seen in the long-slit study of \citet{1998ApJ...502..245G}.  
In the following we describe the main kinematical and morphological features of these three components.

\textit{Linear component.} Over the velocity range 405-745 \kms, at an intensity-weighted mean heliocentric velocity of $\sim 450$ \kms,  we find a well-defined linear, knotty filament. This component is shown in red in the right panel of Fig. \ref{TotalFluxMap} and has a  southwest-northeast orientation, similar to that of the large-scale jet. The average value of the \OIII$\lambda$5007/H$\beta$ ratio is $\sim 6.4^{+0.3}_{-0.3}$ , while for \NII$\lambda$6584/H$\alpha$ it is $\sim 0.56^{+0.09}_{-0.07}$.
An interesting kinematical feature can be seen in one specific location within this component. In Fig.~\ref{pvplot} we show the position-velocity diagram of this feature extracted along the green line drawn in the left panel of Fig.\ref{TotalFluxMap}. Over a well-defined spatial extent (of $\sim$70 pc), the heliocentric velocities of the gas are  300-350 \kms\ higher than those of the surrounding gas (at v$_{\rm hel}\sim$ 450 \kms). The position-velocity diagram shows that the gas gradually reaches higher velocities, creating a continuous velocity structure that gives the impression of a small bubble being accelerated. At this location, the brightness of the filament is markedly lower than in the surrounding regions. 

\textit{Diffuse component.} This component is shown in green in the right panel of Fig. \ref{TotalFluxMap} and has an intensity-weighted mean heliocentric velocity of $\sim$ 260 \kms.  Morphologically, this component appears to be very different from the linear component described above. While some structure on small scales is present, most of the emission comes from extended, diffuse gas that gives the impression of gas filaments being blown to the northeast. Reinforcing this idea, several arc-like clumps with a bow-shock shape can be clearly identified in the southern field.  They vary in size, but all have a semicircular shape and the same orientation.  This diffuse component has a velocity similar, although not identical, to that of the quiescent neutral gas of the nearby \HI\ cloud, and we suggest that it originates from the \HI\ structure.
For the diffuse component the average  \OIII$\lambda$5007/H$\beta$ ratio is $\sim3^{+0.19}_{-0.17}$ while for the \NII$\lambda$6584/H$\alpha$ it is $\sim 0.65^{+0.08}_{-0.07}$.
Similar average values are also found in the arc-like clumps: \OIII$\lambda$5007/H$\beta$ $\sim 4^{+0.25}_{-0.22}$ and \NII$\lambda$6584/H$\alpha$ $\sim 0.64^{+0.07}_{-0.07}$.
The morphology and the velocities of the linear and the diffuse components match - in the overlapping region - those of the high- and low-velocity components observed by \citet{2014arXiv1411.4639S}. 

\textit{Low-velocity component.} At an intensity-weighted mean heliocentric velocity of $\sim 0$ \kms\ , we identify a third component. It is shown in blue in the right panel of Fig. \ref{TotalFluxMap} and has fainter emission than the other two components.  
Morphologically, this component is very similar to the diffuse component described above. It is composed of diffuse emission and, in the lower field, shows a single arc-like structure whose morphology is similar to the arc-like clumps found in the diffuse component.
The average value of the \OIII$\lambda$5007/H$\beta$ ratio is $\sim 1.6 ^{+0.72}_{ -0.64}$, while for the \NII$\lambda$6584/H$\alpha$ ratio it is $\sim 0.81 ^{+0.14}_{-0.11}$.

The average line ratios reported above for the different components are the average of many individual regions extracted for each component. In Fig.~\ref{Spectra} we show a representative spectrum from one of these regions for each component.
We mark the regions from which these spectra are extracted in the left panel of Fig~\ref{TotalFluxMap}.
The average line ratios show that the ionization state of the gas is different in the three  different  components.  
We assume \NII$\lambda$6584/H$\alpha\approx0.6$ and \OIII$\lambda$5007/H$\beta\approx2.0$  as typical dividing values for \HII\ region/AGN and AGN/low-ionization nuclear emission-line region (LINER), respectively \citep[see][]{2006MNRAS.372..961K}. From comparing the extracted line ratios with these values, we confirm earlier results that ruled out that the ionization of the gas is due to stellar light.

\section{Discussion and conclusions}

The MUSE observations reveal the velocity structure of the ionized gas over a large region of the outer filament of Cen~A at high spatial resolution.  The morphology and velocities of the ionized gas in the overlapping region agree with the results of \citet{2014arXiv1411.4639S}. However, the two components that match the kinematics of the components in the nearby \HI\ cloud are now found to cover a much larger area of the filament.
Thus, our data support the results and interpretation from the VIMOS data that the gas is affected by the interaction with the radio jet, and in particular, that this is the case for a much larger area and, by implication, probably for the entire outer filament.
The MUSE data also show that the situation is even more complex. Interestingly, the different kinematical components now also appear to be connected with different morphologies (linear, arc-like, and diffuse).  Furthermore, we identified a new low-velocity component.

\begin{figure}[]
\centering
\includegraphics[width=7 cm, keepaspectratio]{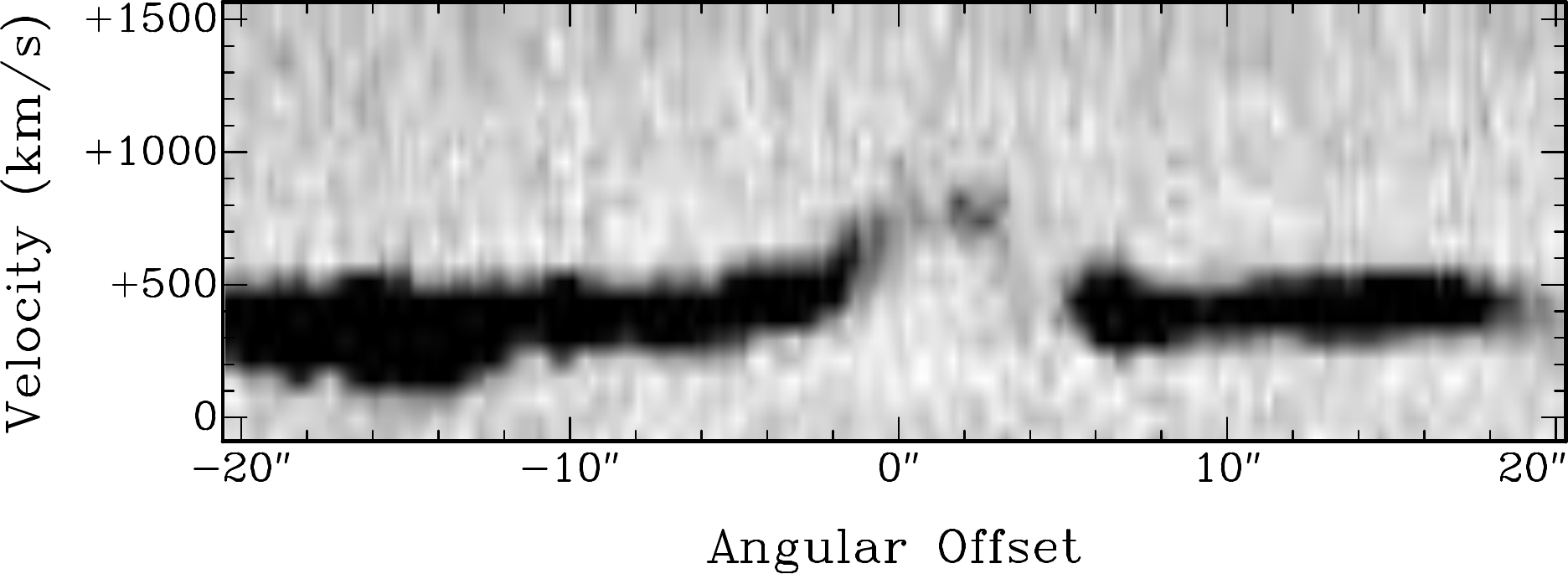}
\caption{Position-velocity plot of the \OIII$\lambda$5007\AA\ emission taken along the line indicated in Fig.~\ref{TotalFluxMap} (left panel). The velocities are heliocentric.}
\label{pvplot}
\end{figure}

For the outer filament we propose that the diffuse and the low-velocity components are directly affected by the passage of a slow-moving jet.  A smooth interaction of this kind would, indeed, explain the arc-like clumps embedded within them. They show a striking similarity with the structures described by \citet{1991PASAu...9...93B}. These simulations show that arc-shaped structures can form as
a result of the encounter between a transonic stream of non-thermal plasma and denser clouds of thermal gas. The orientation of the large-scale jet and the arc-like structures agrees with the results from such simulations.  
In addition, the nature of the large-scale jet of Cen~A is compatible with a low-Mach number jet as used in the simulation.
Consistent with our previous study, we find that the linear component is characterized by velocities similar to the \HI\ {\sl anomalous} velocities, which supports the scenario described in \citet{2014arXiv1411.4639S}. This component represents the part of the \HI\ cloud that through its rotation about the galaxy has entered the zone of influence of the large-scale radio jet. The gas in this region is probably affected by the lateral expansion of the jet cocoon. The well-defined  elongation of the linear component along the same direction as the jet  
is possibly the result of this lateral expansion. These characteristics, and the compression that may result, could also explain the interesting kinematical structure illustrated in Fig.~\ref{pvplot}. These structures are reminiscent of a bursting over-pressured bubble.  
It is unclear whether the newly found low-velocity component is a possible counterpart of this linear component (i.e., also affected by a lateral expansion) but on the other side of the jet and, as a consequence, characterized by low velocities. The very different morphologies of these two components could be due to intrinsic differences in the density or distribution of the gas within the two regions.

\begin{figure}[t]
\centering
\includegraphics[width=8 cm, height=8 cm]{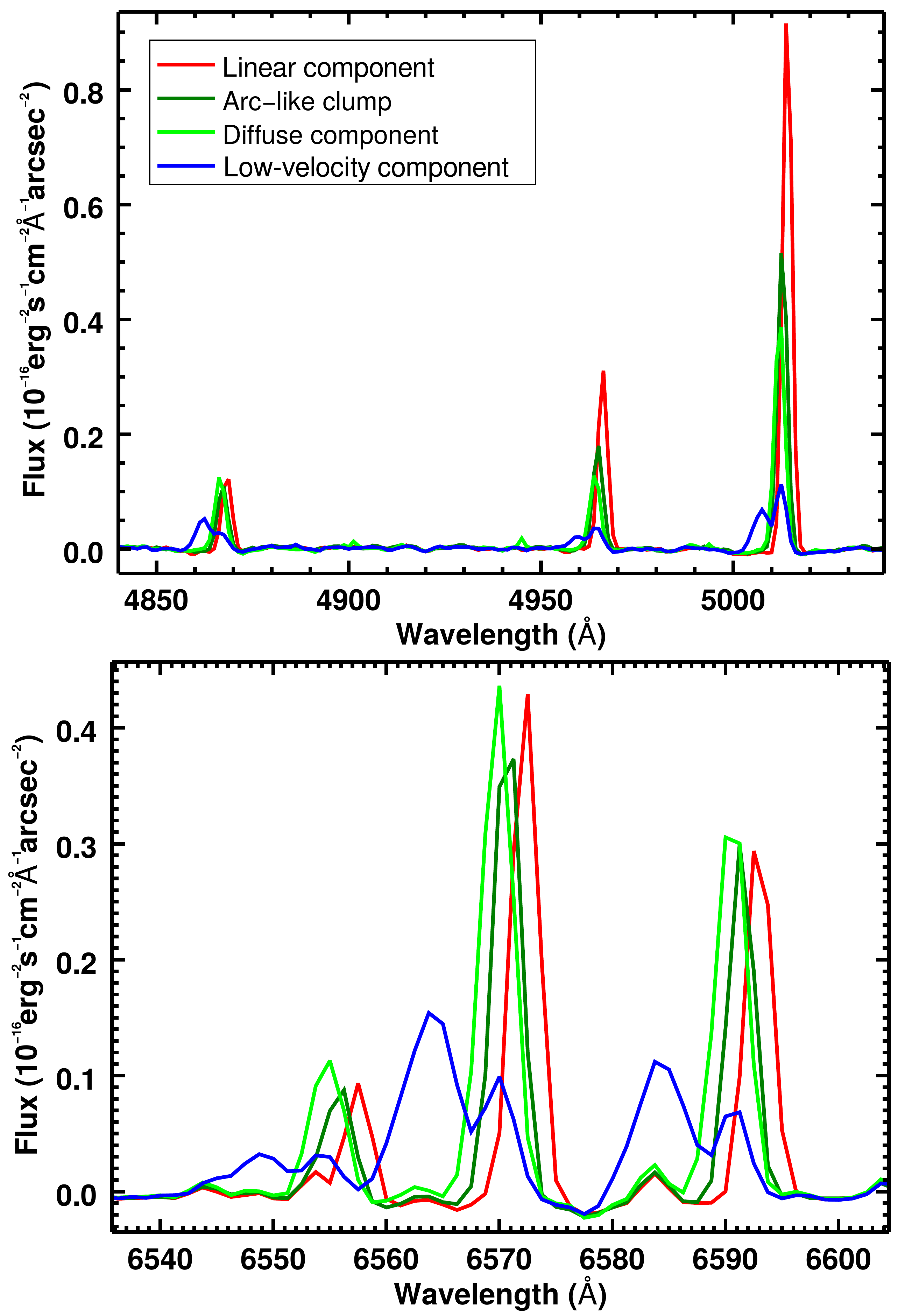}
\caption{Spectra extracted from the regions indicated in Fig.~\ref{TotalFluxMap} (left panel).  
The spectra are given for the H$\beta$, \OIII$\lambda\lambda$4959,5007\AA\ (upper panel) and the H$\alpha$, \NII$\lambda\lambda$6548,6584\AA\ (lower panel) lines. Because it overlaps the diffuse component, the spectrum extracted for the low-velocity component shows double-peaked emission lines. The peak associated with the latter component is at lower wavelengths.}
\label{Spectra}
\end{figure}

Although the line ratios found for all components, when plotted in a BPT diagram \citep{1981PASP...93....5B}, are typical of AGN/LINER, we note a trend to be present, with the linear component showing higher excitation levels.  This may reflect a change in the ionized gas density or in the number of ionizing photons across the region we are observing.  Apart from the low-velocity component, for which the full-width-at-half-maximum (FWHM) of the H$\alpha$ line is $\sim$ 4.5 \AA\ (corresponding to $\sim$ 205 \kms), the narrow velocity width of the H$\alpha$ lines of the other components (FWHM $\sim$ 3~\AA\ corresponding to $\sim$ 137 \kms) leads us to argue \citep[as already discussed in][]{2014arXiv1411.4639S} that the jet does not drive a strong shock across the whole observed region. A more detailed analysis of the kinematics and ionization will be presented in a forthcoming paper.

It is worth noting that we find similarities with the results recently obtained (also using MUSE) for the inner filament of Cen~A \citep{2014arXiv1409.7700H}. Indeed, both the inner and the outer filaments contain a well-defined linear knotty plus a more diffuse structure that show different kinematics and line ratios. It is thus  possible that similar mechanisms are responsible for  the kinematics and the ionization state of the ionized gas within both the inner and the outer filament.

In summary, the general implications of our study are that the kinematics of the gas supports the idea that the large-scale jet is still an active structure, but is characterized, as expected from its morphology,  by relatively low (transonic) velocities. 
Despite being a low-power jet, the effects of its lateral and head-on interaction with the surrounding ISM are clearly visible, and this is the first time that we can trace the gas and its kinematics at the location of a jet-cloud interaction
on such small scales. This shows that the jet-ISM interaction has to be considered when studying the effect of the feedback on a galaxy that hosts a low-power radio source.

\begin{acknowledgements}
GT would like to thank Bernd Husemann for his help with the MUSE pipeline. The research leading to these results has received funding from the European Research Council under the European Union's Seventh Framework Programme (FP/2007-2013) / ERC Advanced Grant RADIOLIFE-320745. Based on observations made with ESO Telescopes at the La Silla Paranal Observatory under programme 60.A-9341A  
\end{acknowledgements}

\bibliographystyle{aa}
\bibliography{biblio.bib}

\end{document}